\input{aipcheck}

\documentclass[
    ,final            
  ]
  {aipproc}

\layoutstyle{8x11double}

\begin{document}

\title{Intracluster Planetary Nebulae as dynamical probes of the diffuse 
light in galaxy clusters}


\author{Magda Arnaboldi}{
  address={INAF, Osservatorio Astronomico di Torino, Strada Osservatorio, 20, 
I-10025 Pino Torinese} }


\begin{abstract}
I will review the latest results for the presence of diffuse light in
the nearby universe and at intermediate redshift, and then discuss the
latest results from hydrodynamical cosmological simulations of cluster
formation on the expected properties of diffuse light in clusters.  I
shall present how intracluster planetary nebulae (ICPNe) can be used
as excellent tracers of the diffuse stellar population in nearby
clusters, and how their number density profile and radial velocity
distribution can provide an observational test for models of cluster
formation. The preliminary comparison of available ICPN samples with
predictions from cosmological simulations support late infall as the
most likely mechanism for the origin of diffuse stellar light in
clusters.
\end{abstract}

\maketitle


\section{Observations of Diffuse Light}

The study of the intracluster light (ICL) began with Zwicky's (1951)
claimed discovery of an excess of light between galaxies in the Coma
Cluster.  Its low surface brightness ($\mu_B > 28$ mag arcsec$^{-2}$)
makes it difficult to study the ICL systematically (Oemler 1973; Thuan
\& Kormendy 1977; Bernstein et al. 1995; Gregg \& West 1998; Gonzalez et
al. 2000). 

Presence of diffuse light can be revealed by tails, arcs or plumes
which are narrow (about $\sim 2$ kpc) and extended ($\sim 50 - 100$
kpc), or as a halo of light at the cluster scales, which is present in the 200
-700 kpc radial range. In the Coma cluster, Adami et al. (2005) have
searched for ICL small features using a wavelet analysis and
reconstruction technique. They identified 4 extended sources, with 50
to 100 kpc diameter and V band magnitudes in the 14.5 - 16.0 range.
Quantitative, large scale measurements of the diffuse light in the
Virgo cluster were recently attempted by Mihos et al. (2005): these
deep observations show the intricate and complex structure of the ICL in
Virgo.

On the cluster scales, the presence of diffuse light can be revealed
when the whole distribution of stars in clusters is analysed in a way
similar to Schombert's (1986) photometry of brightest cluster galaxies
(BCGs). When this component is present, the surface-brightness
profiles centred on the BCG turn strongly upward in a
$(\mu,R^{1/\alpha})$ plot for radii from 200 to 700 kpc.  This
approach to ICL low surface brightness measurements was taken by
Zibetti et al. (2005), who studied the spatial distribution and colors
of the ICL in 683 clusters of galaxies at $z\simeq 0.25$ by stacking
their images, after rescaling them to the same metric size and masking
out resolved sources.

In nearby galaxy clusters, intracluster planetary nebulae (ICPNe) can
be used as tracers of the ICL; this has the advantages that detection
of ICPNe are possible with deep narrow band images and that the ICPN
radial velocities can be measured to investigate the dynamics of the
ICL component. ICPN candidates have been identified in Virgo
(Arnaboldi et al. 1996, 2002, 2003; Feldmeier et al. 2003,
2004a) and Fornax (Theuns \& Warren 1997), with significant numbers of
ICPN velocities beginning to become available (Arnaboldi et al. 2004).
 
The overall amount of the ICL in galaxy clusters is still a matter of
debate. However, there is now observational evidence that it may
depend on the physical parameters of clusters, with rich galaxy
clusters containing 20\% or more of their stars in the intracluster
component (Gonzalez et al. 2000; Gal-Yam et al. 2003), while the Virgo
Cluster has a fraction of $\sim 10$\% in the ICL (Ferguson et
al. 1998; Durrell et al. 2002; Arnaboldi et al. 2002, 2003; Feldmeier
et al. 2004a), and the fraction of detected intragroup light (IGL) is
1.3\% in the M81 group (Feldmeier et al. 2004b) and less than 1.6\% in
the Leo I group (Castro-Rodr\'iguez et al. 2003). Recent
hydrodynamical simulations of galaxy cluster formation in a
$\Lambda$CDM cosmology have corroborated this observational evidence:
in these simulated clusters, the fraction of the ICL increases from
$\sim$ 10\% $-$ 20\% in clusters with $10^{14} M_\odot$ to up to 50\%
for very massive clusters with $10^{15} M_\odot$ (Murante et
al. 2004). Strong correlation between ICL fraction and cluster mass is
also predicted from semi-analytical models of structure formation
(Lin \& Mohr 2004).

The mass fraction and physical properties of the ICL and their
dependence on cluster mass will be related with the mechanisms by
which the ICL is formed. Theoretical studies predict that if most of
the ICL is removed from galaxies because of their interaction with the
galaxy cluster potential or in fast encounters with other galaxies,
the amount of the ICL should be a function of the galaxy number
density (Richstone \& Malumuth 1983; Moore et al. 1996). The early
theoretical studies about the origin and evolution of the ICL
suggested that it might account for between 10\% and 70\% of the total
cluster luminosity (Richstone \& Malumuth 1983; Malumuth \& Richstone
1984; Miller 1983; Merritt 1983, 1984). These studies were based on
analytic estimates of tidal stripping or simulations of individual
galaxies orbiting in a smooth gravitational potential. Nowadays,
cosmological simulations allow us to study in detail the evolution of
galaxies in cluster environments (see, e.g., Moore et al. 1996;
Dubinski 1998; Murante et al. 2004; Willman et al. 2004; Sommer-Larsen
et al. 2005). Napolitano et al. (2003) investigated the ICL for a
Virgo-like cluster in one of these hierarchical simulations,
predicting that the ICL in such clusters should be unrelaxed in
velocity space and show significant substructures; spatial
substructures have been observed in one field in the ICPNe identified
with [O III] and H$\alpha$ (Okamura et al. 2002).

\section{Diffuse light in clusters from cosmological simulations}

Cosmological simulations of structure formation facilitate studies of
the diffuse light and its expected properties.  Dubinski (1998)
constructed compound models of disk galaxies and placed them into a
partially evolved simulation of cluster formation, allowing an
evolutionary study of the dark matter and stellar components
independently.  Using an empirical method to identify stellar tracer
particles in high-resolution cold dark matter (CDM) simulations, Napolitano
et al. (2003) studied a Virgo-like cluster, finding evidence of a
young dynamical age of the intracluster component.  The main
limitations in these approaches is the restriction to collisionless
dynamics.

Murante et al. (2004) analyzed for the first time the ICL formed in
a cosmological hydrodynamical simulation including a self-consistent
model for star formation. In this method, no assumptions about the
structural properties of the forming galaxies need to be made, and the
gradual formation process of the stars, as well as their subsequent
dynamical evolution in the non-linearly evolving gravitational
potential can be seen as a direct consequence of the $\Lambda$CDM
initial conditions.  

Murante et al. (2004) identified 117 clusters in
a large volume of $192^3\, h^{-3}{\rm Mpc}^3$, and analyze the
correlations of properties of diffuse light with, e.g., cluster mass
and X-ray temperatures. Galaxies at the centers of these
clusters have surface-brightness profiles which turn strongly upward
in a $(\mu,R^{1/\alpha})$ plot. This light excess can be explained as
IC stars orbiting in the cluster potential.  Integrating its density
distribution along the line-of-sight (LOS), the slopes from Murante et
al. (2004) simulations are in agreement with those observed for the
surface brightness profiles of the diffuse light in nearby clusters.

\begin{figure}
  \includegraphics[height=.3\textheight]{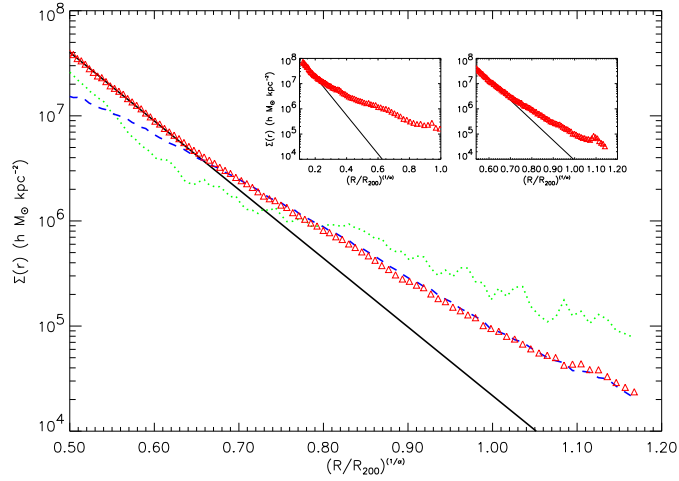}
  \caption{Schombert--like analysis on the {\it stacked} 2D radial
density profile (BCG + ICL) of clusters in
the Murante et al (2004) simulation (triangles). The light excess is
evident at large cluster radii. The solid line shows the function $
\log \Sigma(r) = \log \Sigma_e -3.33 [(r/r_e)^{1/\alpha} -1]$, with
best--fit parameters $\log \Sigma_e = 20.80$, $r_e = 0.005$, $\alpha =
3.66$ to the BCG inner stellar light. Also shown are the averaged 2D density
profile of stars in galaxies (dotted line) and in the field (dashed
line).  In the inserts, the results are shown from the same analysis
for the most luminous clusters with $T>4$ keV (left panel), and for
less luminous ones with $0<T<2$ keV (right panel).  The resulting
best--fit parameters are respectively $\log \Sigma_e = 16.47$, $r_e =
0.11$, $\alpha = 1.24$ and $\log \Sigma_e = 23.11$, $r_e = 0.00076$,
$\alpha = 4.37$.  In the main plot and in the inserts the unit
$(R/R_{200})^{1/\alpha}$ refers to the $\alpha$ values given by each
Sersic profile. From Murante et al. (2004).}\label{fig1}
\end{figure}

At large cluster radii, the surface brightness profile of the ICL
appears more centrally concentrated than the surface brightness
profile of cluster galaxies (see Figure~\ref{fig1} ). The prediction of
ICL being more centrally concentrated than the galaxy cluster light
has been tested observationally.  Zibetti et al. (2005) have presented
surface photometry from the stacking of 683 clusters of galaxies
imaged in the g- , r-, and i-bands in the SDSS. They have been able to
measure surface brightness as deep as $\mu_r \sim 32$ mag
arcsec$^{-2}$ for the ICL light and $\mu_r \sim 29.0$ for the total
light, out to 700 kpc from the BCG.  They finds that the ICL is
significantly more concentrated than the total light.

\begin{figure}
\includegraphics[height=.3\textheight]{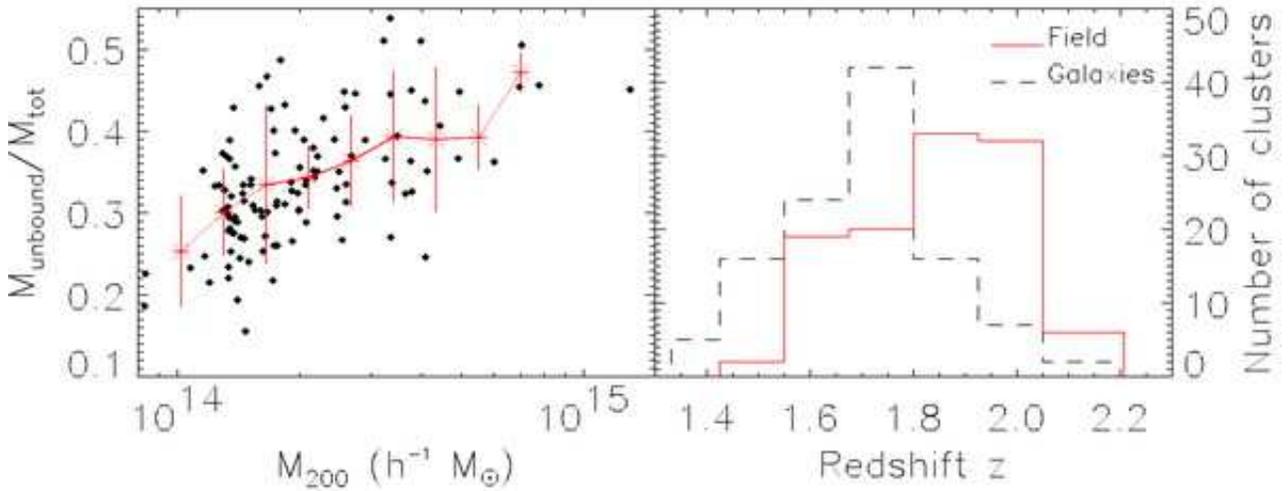}
\caption{Left: Fraction of stellar mass in diffuse light vs.\ cluster
mass. Dots are for clusters in the simulated volume; asterisks show
the average values of this fraction in 9 mass bins with
errorbars. Right panel: histograms of clusters over mean formation
redshift, of their respective bound (dashed) and IC star particles
(solid line). Mean formation redshifts are evaluated for each cluster as
the average on the formation redshift of each star particle. From
Murante et al. (2004).}
\label{fig2}
\end{figure}

From the simulations carried out by Murante et al. (2004), they also
obtained the redshifts $z_{form}$ at which the stars formed: those in
the IC component have a $z_{form}$ distribution which differs from
that in cluster galaxies, see Figure~\ref{fig2}. The ``unbound'' stars
are formed earlier than the stars in galaxies. The prediction for an
old stars' age in the diffuse component agrees with the HST
observation of the IRGB stars in the Virgo IC field, e.g. $t> 2$Gyr
(Durrell et al. 2002), and points toward the early tidal interactions
as the preferred formation process for the ICL.  The different age and
spatial distribution of the stars in the diffuse component indicate
that it is a stellar population that is not a random sampling of the
stellar populations in cluster galaxies.

Murante et al. (2004) studied the correlation between the fraction of
stellar mass in the diffuse component and the clusters' total mass in
stars, based on their statistical sample of 117 clusters.  This
fraction is $\sim 0.1$ for cluster masses $M > 10^{14}h^{-1}M_\odot$
and it increases with cluster mass: the more massive clusters have the
largest fraction of diffuse light, see Figure~\ref{fig2}). For $M \sim
10^{15}h^{-1}M_\odot$, simulations predict as many stars in the
diffuse component as in cluster galaxies.

\subsection{Predicted dynamics of the ICL}
In the currently favored hierarchical clustering scenario, fast
encounters and tidal interactions within the cluster potential are the
main players of the morphological evolution of galaxies in clusters.
Fast encounters and tidal stirring cause a significant fraction of the
stellar component in individual galaxies to be stripped and dispersed
within the cluster in a few dynamic times. If the timescale for
significant phase-mixing is on the order of few cluster internal
dynamical times, then a fraction of the ICL should still be located in
long streams along the orbits of the parent galaxies. Detections of
substructures in phase space would be a clear sign of late infall and
harassment as the origin of the ICL.

A high resolution simulation of a Virgo-like cluster in a $\Lambda
CDM$ cosmology was used to predict the velocity and the clustering
properties of the diffuse stellar component in the intracluster region
at the present epoch (Napolitano et al. 2003). The simulated cluster
builds up hierarchically and tidal interactions between member
galaxies and the cluster potential produce a diffuse stellar component
free-flying in the intracluster medium. The simulations are able to
predict the radial velocity distribution expected in spectroscopic
follow-up surveys: they find that at $z=0$ the intracluster stellar
light is mostly dynamically unmixed and clustered in structures on
scales of about 50 kpc at a radius of $400-500$ kpc from the cluster
center.

\begin{figure}
\includegraphics[height=.3\textheight]{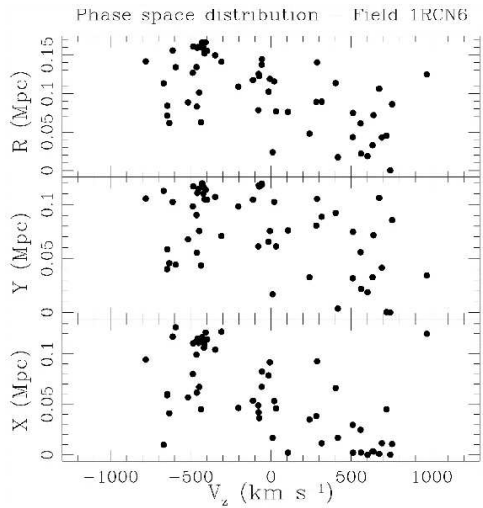}
\caption{Projected phase-space diagram for a simulated ICPN sample in a
  N-body simulation of a Virgo-like cluster. From Napolitano et al. (2003).}
\label{fig3}
\end{figure}

Willman et al. (2004) and Sommer-Larsen et al. (2005) have studied the
dynamics of the ICL in cosmological hydrodynamical simulations for the
formation of a rich galaxy cluster. In a Coma-like rich cluster,
Willman et al. (2004) finds that the ICL show significant substructure
in velocity space, tracing separate streams of stripped IC
stars. Evidence is given that despite an un-relaxed distribution, IC
stars are useful mass tracers, when several fields at a range of radii
have measured LOS velocities. According to
Sommer-Larsen et al. (2005), IC stars are colder than cluster
galaxies. This is to be expected because diffuse light is more
centrally concentrated than cluster galaxies, as found in cosmological
simulations (see Murante et al. 2004) and confirmed from observations
of intermediate redshift clusters (Zibetti et al. 2005), and both the
ICL and galaxies are in equilibrium with the same cluster potential.

\section{Intracluster planetary nebulae in the Virgo cluster: the projected 
phase space distribution}

Intracluster planetary nebulae (ICPNe) have several unique features
that make them ideal for probing the ICL. The diffuse
envelope of a PN re-emits 15\% of the UV light of the central star in
one bright optical emission line, the green [OIII]$\lambda 5007$ \AA\
line. PNe can therefore readily be detected in external galaxies out
to distances of 25 Mpc and their velocities can be determined from
moderate resolution $(\lambda /\Delta \lambda \sim 5000)$ spectra:
this enables kinematical studies of the IC stellar population.

PNe trace stellar luminosity and therefore provide an estimate of
the total IC light. Also, through the [OIII] $\lambda 5007$ \AA\
planetary nebulae luminosity function (PNLF), PNe are good distance
indicators, and the observed shape of the PNLF provides information on
the LOS distribution of the IC starlight. Therefore
ICPNe are useful tracers to study the spatial distribution, kinematics,
and metallicity of the diffuse stellar population in nearby clusters.

\subsection{Current narrow band imaging surveys}
Several groups (Arnaboldi et al. 2002, 2003; Aguerri et al. 2005;
Feldmeier et al. 2003, 2004a) have embarked on narrow-band [OIII]
imaging surveys in the Virgo cluster, with the aim of determining the
radial density profile of the diffuse light, and gaining information
on the velocity distribution via subsequent spectroscopic observations
of the obtained samples. Given the use of the PNLF as distance
indicators, one also obtain valuable information on the 3D shape of the
Virgo cluster from these ICPN samples.

\begin{figure}
\includegraphics[width=.6\textwidth]{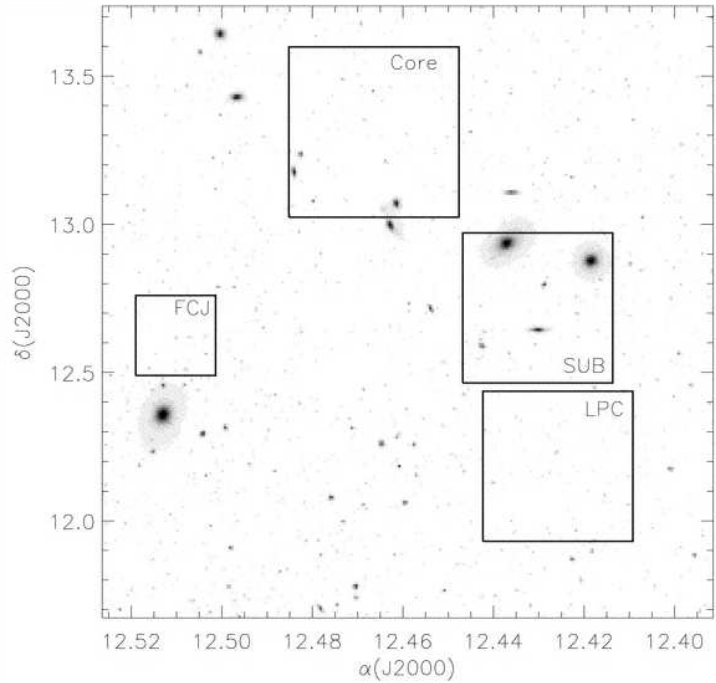}
\caption{Aguerri et al. (2005) surveyed fields in the Virgo cluster
core. The CORE field was obtained at the ESO MPI 2.2m telescope, and
the SUB field with the Suprime Cam at the 8.2m Subaru telescope. The
FCJ field is from prime focus camera of the Kitt Peak 4m telescope and
the lower right field, LPC, was observed at the La Palma INT. From
Aguerri et al. (2005).}
\label{fig4}
\end{figure}

Wide-field mosaic cameras, such as the WFI on the ESO MPI 2.2m
telescope and the Suprime Cam on the Subaru 8.2m, allow us to identify
the ICPNe associated with the extended ICL (Arnaboldi et al. 2002,
2003; Okamura et al. 2002; Aguerri et al. 2005). These surveys require
the use of data reduction techniques suited for mosaic images, and
also the development and refining of selection criteria based on
color-magnitude diagrams from photometric catalogs, produced with
SExtractor (Bertin \& Arnout 1996).

\begin{figure}
\includegraphics[height=.3\textheight]{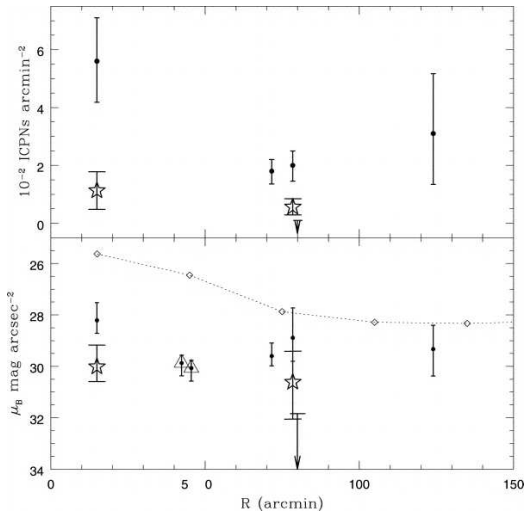}
\caption{Number density of PNe (top) and surface brightness (bottom) in our 
surveyed fields. In the top panel, circles show the measured number
densities from Table 3 of Aguerri et al. (2005), and error bars denote
the Poisson errors. For the LPC field our upper limit is given. For
the RCN1 field at the largest distance from M87, the uncertainty from
the correction for Ly$\alpha$ emitters is substantial and is included
in the error bar. The large stars with Poisson error bars show the
number densities of PNe in FCJ and SUB fields not including PNe bound
to the halos of M87 and M84. In the lower panel, circles show the
surface brightness inferred with the average value of $\alpha$ in
Table 4 of Aguerri et al. (2005), and error bars show the range of
values implied by the Poisson errors and the range of adopted $\alpha$
values. Triangles represent the measurements of the ICL from RGB
stars; error bars indicate uncertainties in the metallicity, age, and
distance of the parent population as discussed in Durrell et
al. (2002). The stars indicate the surface brightness associated with
the ICPNe in the FCJ and SUB fields that are not associated with the
M87 or M84 halos but are free flying in the Virgo Cluster potential,
(Arnaboldi et al. 2004). The dashed line and diamonds show the B-band
luminosity of Virgo galaxies averaged in rings (Binggeli et
al. 1987). Distances are relative to M87. The ICL shows no trend with
cluster radius out to 150 arcmin. From Aguerri et al. (2005).}
\label{fig6}
\end{figure}

The data analysed by Aguerri et al. (2005) constitute a sizable
sample of ICPNe in the Virgo core region, constructed homogeneously
and according to rigorous selection criteria; a layout of the
pointings is shown in Figure~\ref{fig4}. From the study of five
wide-fields they conclude that the number density plot in 
Figure~\ref{fig6} shows no clear trend with distance from the cluster
center at M87, except that the value in the innermost FCJ field is
high.  However, the spectroscopic results of Arnaboldi et al. (2004)
have shown that 12/15 PNe in this field have a low velocity dispersion
of 250 kms$^{-1}$, i.e. in fact they belong to the outer halo of M87,
which thus extends to at least 65 kpc radius. In the SUB field, 8/13
PNe belong to the similarly cold, extended halo of M84, while the
remaining PNe are observed at velocities that are close to the
systemic velocities of M86 and NGC 4388, the two other large galaxies
in or near this field. It is possible that in a cluster as young and
unrelaxed as Virgo, a substantial fraction of the ICL is still bound
to the extended halos of galaxies, whereas in denser and older
clusters these halos might already have been stripped. If so, it is
not inappropriate to already count the luminosity in these halos as
part of the ICL. However, in Figure~\ref{fig6} the plot of the PN
number density with radius is also shown for the case in which the PNe
in the outer halos of M87 and M84 are removed from the FCJ and SUB
samples. In this case, the resulting number density is even more
nearly flat with radius, but there are still significant
field-to-field variations; in particular, the remaining number
densities in SUB and LPC are low.

When one wishes to compare the luminosity of the ICL at the positions
of Aguerri et al. (2005) fields with the luminosity from the Virgo
galaxies, one adds in further uncertainties, because the luminosities
of nearby Virgo galaxies depend very much on the location and field
size surveyed in the Virgo Cluster. Aguerri et al. (2005) consider
therefore their reported intervals in surface brightness to be their
primary result, while the relative fractions of the ICL with respect
to the Virgo galaxy light are evaluated for comparison with previous
ICPN works, and considered them to be more uncertain.

From the study of four wide fields in the Virgo core, Aguerri et
al. (2005) obtain a mean surface luminosity density of $2.7 \times
10^6$ L$_{B\odot}$ arcmin$^{-2}$, rms = $2.1 \times 10^6$ L$_{B\odot}$
arcmin$^{-2}$, and a mean surface brightness of $\mu_B$ = 29.0 mag
arcsec$^{-2}$. Their best estimate of the ICL
fractions with respect to light in galaxies in the Virgo core is $\sim
5\%$. However, there are significant field-to-field variations. The
fraction of the ICL versus total light ranges from $\sim 8\%$ in the
CORE and FCJ fields, to less than 1\% in the LPC field, which in its
low ICL fraction is similar to low-density environments
(Castro-Rodr\'iguez et al. 2003). This latter field corresponds to the lowest
luminosity density in the mosaic image of the Virgo core region from
Mihos et al. (2005).

\section{Spectroscopic follow-up}
ICPNe are the only component of the ICL whose kinematics can be
measured at this time. This is important since the high-resolution
N-body and hydrodynamical simulations predict that the ICL is
un-relaxed, showing significant substructure in its spatial and
velocity distributions in clusters similar to Virgo. 

The spectroscopic follow-up with FLAMES of the ICPN candidates
selected from three survey fields in the Virgo cluster core was
carried out by Arnaboldi et al. (2004).  Radial velocities of 40 ICPNe
in the Virgo cluster were obtained with the new multi-fiber FLAMES
spectrograph on UT2 at VLT.  The spectra were taken for a
homogeneously selected sample of ICPNe, previously identified in three
$\sim 0.25$ deg$^2$ fields in the Virgo cluster core.  For the first
time, the $\lambda$ 4959 \AA\ line of the [OIII] doublet is seen in a
large fraction (40\%) of ICPNe spectra, and a large fraction of the
photometric candidates with m(5007) $ < 27.2$ is spectroscopically
confirmed.

\subsection{The LOS velocity distributions of ICPNe in the
Virgo cluster core.}  
\begin{figure}
\includegraphics[height=.6\textheight,angle=-90]{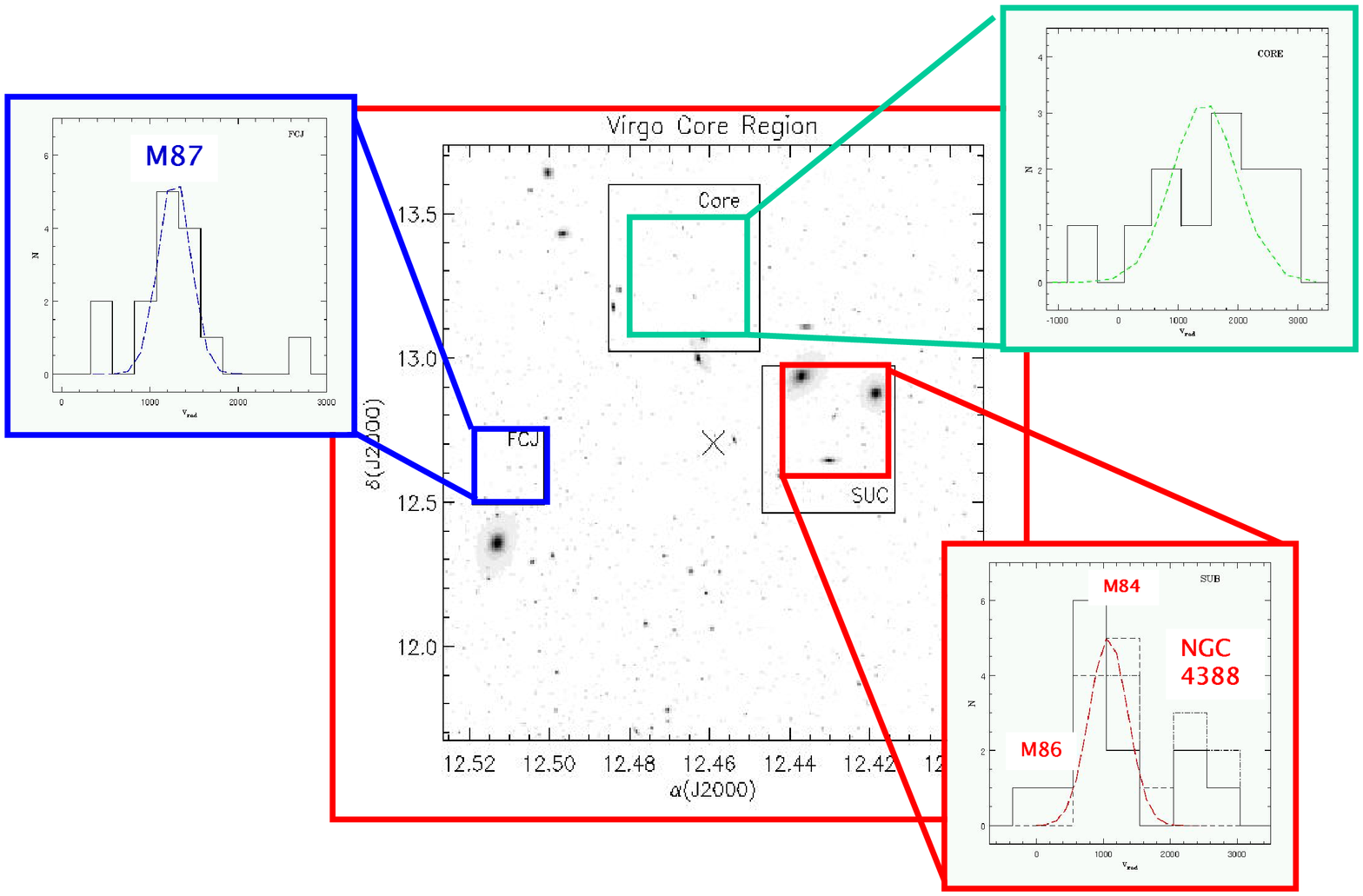}
\caption{ICPN radial velocity distributions in the three pointings
(FCJ, CORE, and SUB) from Aguerri et al. (2005). In FCJ panel, blue
dashed line shows a Gaussian with $\bar{v}_{rad} = 1276$ km s$^{-1}$
and $\sigma_{rad} = 247$ km s$^{-1}$. In CORE, green dashed line shows
a Gaussian with $\bar{v}_{rad} = 1436$ km s$^{-1}$ and $\sigma_{rad} =
538$ km s$^{-1}$, for VC galaxies dE and dS0 within 2$^\circ$ of M87
(from Binggeli et al. 1987). In SUB, dashed histogram shows radial
velocities from TNG spectroscopic follow-up (Arnaboldi et
al. 2003). Dashed red line shows a Gaussian with $\bar{v}_{rad} =
1080$ km s$^{-1}$ and $\sigma_{rad} = 286$ km s$^{-1}$. Dashed-dotted
lines show the SUB-FLAMES spectra including those spectra for HII
regions, which have radial velocities in M84 \& NGC~4388 redshift
ranges.}
\label{fig5}
\end{figure}

With these data, Arnaboldi et al. (2004) were able for the first time
to determine radial velocity distributions of ICPNe and use these to
investigate the dynamical state of the Virgo cluster. Figure~\ref{fig5}
shows an image of the Virgo cluster core with the positions of the
imaged fields. The radial velocity distributions obtained from the
FLAMES spectra in three of these fields are also displayed in
Figure~\ref{fig5}.  Clearly the velocity distribution histograms for the
three pointings are very different.

In the FCJ field, the ICPNe distribution is dominated by the halo of
M87. There are 3 additional outliers, 2 at low velocity, which are
also in the brightest PNLF bin, and therefore may be in front of the
cluster.  The surface brightness of the ICL associated with
the 3 outliers, e.g.\ the ICPNe in the FCJ field, amounts to $\mu_B \simeq
30.63$ mag arcsec$^{-2}$, in agreement with the surface brightness
measurements of Ferguson et al. (1998) and Durrell et al. (2002) of
the intracluster red giant stars.

The M87 peak of the FCJ velocity distribution contains 12 velocities
with $\bar{v}_{p} = 1276\pm 71$ km s$^{-1}$ and $\sigma_{p} = 247\pm
52$ km s$^{-1}$. The average velocity is consistent with that of M87,
$v_{sys} = 1258$ km s$^{-1}$. The distance of the center of the FCJ
field from the center of M87 is $15.'0\simeq\,65$ kpc for an assumed
M87 distance of $15$ Mpc. The value of $\sigma_p$ is very consistent
with the stellar velocity dispersion profile extrapolated outwards
from $\simeq 150''$ in Figure~5 of Romanowsky \& Kochanek (2001) and
falls in the range spanned by their dynamical models for the M87
stars.  The main result from our measurement of $\sigma_p$ is that M87
has a stellar halo in approximate dynamical equilibrium out to at
least $65$ kpc.

In the CORE field, the distribution of ICPN LOS velocities is clearly
broader than in the FCJ field. It has $\bar{v}_{C} = 1491\pm 290$ km
s$^{-1}$ and $\sigma_{C} = 1000\pm 210$ km s$^{-1}$.  The CORE field
is in a region of Virgo devoid of bright galaxies, but contains 7
dwarfs, and 3 low luminosity E/S near its S/W borders. None of the
confirmed ICPNe lies within a circle of three times half the major
axis diameter of any of these galaxies, and there are no correlations
of their velocities with the velocities of the nearest galaxies where
these are known. Thus in this field there is a true IC stellar
component.

The mean velocity of the ICPN in this field is consistent with that of
25 Virgo dE and dS0 within 2$^\circ$ of M87, $<v_{\rm dE,M87}> =
1436\pm108$ km s$^{-1}$ (Binggeli et al.\ 1987), and with that of 93
dE and dS0 Virgo members, $<v_{\rm dE,Virgo}> = 1139\pm67$ km s$^{-1}$
(Binggeli et al.\ 1993).  However, the velocity dispersion of these
galaxies is smaller, $\sigma_{\rm dE,M87}=538\pm 77$ km s$^{-1}$ and
$\sigma_{\rm dE,Virgo}=649\pm 48$ km s$^{-1}$.

The inferred luminosity from the ICPNe in the CORE field is $1.8\times
10^9 L_{B,\odot}$. This is about three times the luminosity of all
dwarf galaxies in this field, $5.3\times 10^8 L_{B,\odot}$, but an
order of magnitude less than the luminosities of the three
low-luminosity E/S galaxies near the field borders. Using the results
of Nulsen \& B\"ohringer (1995) and Matsushita et al.\ (2002),
Arnaboldi et al. (2004) estimate the mass of the M87 subcluster inside
310 kpc (the projected distance $D$ of the CORE field from M87) as
$4.2\times 10^{13} M_\odot$, and compute a tidal parameter $T$ for all
these galaxies as the ratio of the mean density within the CORE field
to the mean density of the galaxy. They find $T=0.01-0.06$, independent
of galaxy luminosity.  Since $T\sim D^{-2}$, any of these galaxies
whose orbit {\sl now} comes closer to M87 than $\sim 60$ kpc would be
subject to severe tidal mass loss.  Based on the
evidence so far, a tantalizing possibility is that the ICPN population
in the CORE field could be debris from the tidal disruption of small
galaxies on nearby orbits in the M87 halo. 

In the SUB field the velocity distribution from FLAMES spectra is
again different from CORE and FCJ. The histogram of the LOS velocities
shows substructures related to M86, M84 and NGC 4388, respectively,
and in Figure~\ref{fig5} the projected phase space is shown.  The
association with the three galaxies is strengthened when we plot the
LOS velocities of 4 HII regions (see Gerhard et al.\ 2002) detected
with FLAMES in this pointing. The substructures in this distribution
are highly correlated with the galaxy systemic velocities.  The
highest peak in the distribution coincides with M84, and even more so
when we add the LOS velocities obtained previously at the TNG
(Arnaboldi et al.\ 2003).  The 10 TNG velocities give $\bar{v}_{\rm
M84} = 1079\pm 103$ km s$^{-1}$ and $\sigma_{\rm M84} = 325\pm75$ km
s$^{-1}$ within a square of $4 R_e \times 4 R_e$ of the M84 center.
The 8 FLAMES velocities give $\bar{v}_{\rm M84} = 891\pm 74$ km
s$^{-1}$ and $\sigma_{\rm M84} = 208\pm54$ km s$^{-1}$, going out to
larger radii.  Note that this includes the over-luminous PNe not
attributed to M84 previously. The combined sample of 18 velocities
gives $\bar{v}_{\rm M84} = 996\pm 69$ km s$^{-1}$ and $\sigma_{\rm
M84} = 293\pm50$ km s$^{-1}$.  Most likely, all these PNe belong to a
very extended halo around M84 (see the deep image in Arnaboldi et al.\
1996). It is possible that the somewhat low velocity with respect to
M84 may be a sign of tidal stripping by M86.

\section{Future prospects and Conclusions}

The observations indicate that the diffuse light is important in
understanding cluster evolution, the star formation history and the
enrichment of the Intracluster Medium. Measuring the projected phase
space distribution of the IC stars constrains how and when this light
originates, and the ICPNe are the only abundant stellar component of
the ICL whose kinematics can be measured at this time.

These measurements are not restricted only to clusters within 25 Mpc
distance: by using a technique similar to those adopted for studies of
Ly$\alpha$ emitting galaxies at very high redshift, Gerhard et
al. (2005) were able to detect PNe associated with the diffuse light
in the Coma cluster, at 100 Mpc distance, in a field which was
previously studied by Bernstein et al. (1995); see also O. Gerhard's
contribution, this conference.  Now it has become possible to study
ICL kinematics also in denser environments like the Coma cluster, and
we can explore the effect of environments with different densities on
galaxy evolution.

\begin{theacknowledgments}
M.A. would like to thank the organizing committee of the Conference on
Planetary Nebulae as Astronomical Tools (Gdansk Poland, 28 June-2 July
2005) for the invitation to give this review. This work has been done
in collaboration with Ortwin E. Gerhard, Kenneth C. Freeman, and
J. Alfonso Aguerri, Massimo Capaccioli, Nieves Castro-Rodriguez, John
Feldmeier, Fabio Governato, Rolf-P. Kudritzki, Roberto Mendez,
Giuseppe Murante, Nicola R. Napolitano, Sadanori Okamura, Maurilio
Pannella, Naoki Yasuda. M.A. wishes to thank ESO for the support of
this project and the observing time allocated both at La Silla and
Paranal Telescopes. M.A. wishes to thank the National Astronomical
Observatory of Japan, for the observing time allocated at the Subaru
Telescope.  This work has been supported by INAF and the Swiss
National Foundation.
\end{theacknowledgments}

\end{document}